\documentclass[12pt]{iopart}
\usepackage{graphicx}

\begin{document}

\review[Multiferroicity due to charge ordering]
{Multiferroicity due to charge ordering}

\author{Jeroen van den Brink$^{1,2}$ and Daniel I. Khomskii$^3$}

\address{
$^1$Institute Lorentz for Theoretical Physics, Leiden University, 
          P.O. Box 9506, 2300 RA Leiden, The Netherlands\\ 
$^2$Institute for Molecules and Materials, Radboud Universiteit Nijmegen,
P.O. Box 9010, 6500 GL Nijmegen, The Netherlands\\
$^3$II Physikalisches Institut, Universit\"at zu K\"oln, 50937 K\"oln, Germany}
\ead{brink@lorentz.leidenuniv.nl khomskii@ph2.uni-koeln.de}
\begin{abstract}
In this contribution to the special issue on multiferroics we focus on multiferroicity driven by different forms of charge ordering. 
We will present the generic mechanisms by which charge ordering can induce ferroelectricity in magnetic systems. There is a number of specific classes of materials for which this is relevant. We will discuss in some detail $(i)$ perovskite manganites of the type (PrCa)MnO$_3$, $(ii)$ the complex and interesting situation in magnetite Fe$_3$O$_4$, $(iii)$ strongly ferroelectric frustrated LuFe$_2$O$_4$, $(iv)$ an example of a quasi one-dimensional organic system. All these are ``type-I'' multiferroics, in which ferroelectricity and magnetism have different origin and occur at different temperatures. In the second part of this article we discuss ``type-II'' multiferroics, in which ferroelectricity is completely {\it due to} magnetism, but with charge ordering playing important role, such as $(v)$ the newly-discovered multiferroic Ca$_3$CoMnO$_6$, $(vi)$ possible ferroelectricity in rare earth perovskite nickelates of the type RNiO$_3$, $(vii)$ multiferroic properties of manganites of the type RMn$_2$O$_5$,  $(viii)$ of perovskite manganites with magnetic E-type ordering and $(ix)$ of bilayer manganites. 
\end{abstract}


\maketitle

\section{Introduction} 

Recently there has been a new surge of interest in multiferroics (MF's), single phase compounds in which magnetism and ferroelectrity coexist~\cite{fiebig05,mostovoy07}. Such materials are relatively rare, which raises the fundamental question what possible mechanisms for multiferroic behavior can exist~\cite{hill00,khomskii06}. Here we consider multiferroicity that is driven by different forms of charge ordering.

\subsection{Type-I and Type-II multiferroics}
Crudely one can divide multiferroics into two groups~\cite{remark_homogeneous}. In  type-I multiferroics ferroelectricity (FE) and magnetism have different origin  and are often due to different active "subsystems" of a material. In such type-I multiferroics the magnetic order parameter, breaking time reversal symmetry, and the ferroelectric order parameter, breaking spatial inversion symmetry, coexist and have a certain coupling between them. 

In the materials that belong to the class of type-I multiferroics FE can have a number of possible microscopic origins~\cite{khomskii06}. A cause of ferroelectricity can be ($i$) the presence of transition metal (TM) with d$^0$ configuration, just as in BaTiO$_3$; ($ii$) the presence of bismuth or lead where the FE is predominantly due to lone pairs of Bi$^{3+}$ and Pb$^{2+}$; or ($iii$) the presence of "geometric" ferroelectricity as in YMnO$_3$, where FE is caused by a rotation of rigid M-O polyhedra (in this case MnO$_5$ trigonal bi-piramids). In general in these type-I materials the FE ordering temperature is much higher than the magnetic one.

In type-II multiferroics ferroelectricity occurs {\it only} in the magnetically ordered state: ferroelectricity sets in at the same temperature as certain type of magnetic ordering and is driven by it. Spiral magnetic ordering, for example, can give rise to type-II multiferroicity~\cite{nagaosa,mostovoy06}.

A special group of multiferroic are materials in which ferroelectricity is caused by the charge ordering (CO); these systems are the focus of the present paper. Most of the MF's and potential MF's that we will discuss are of type-I: the perovskite manganites (PrCa)MnO$_3$~\cite{efremov04,rivadulla, mercone}, magnetite Fe$_3$O$_4$~\cite{rado75,rado77,miyamoto88,miyamoto93,miyamoto94}, quasi one-dimensional organics, and the frustrated charge ordered system LuFe$_2$O$_4$~\cite{ikeda00}. The complex manganites RMn$_2$O$_5$~\cite{hur04} are probably the first example of  type-II multiferroics in which FE is due to the simultaneous presence of sites with different charges and with inequivalent bonds occurring in a magnetically ordered state. Another such example is the recently discovered MF in Ca$_3$CoMnO$_6$. We will discuss the observation that the combination of CO and magnetism may also cause MF behavior in rare earth nickelates RNiO$_3$ and point out that related physics explains MF behavior of perovskite manganites with magnetic E-type ordering as well as of particular bilayer manganites with CO .

\subsection{How charge ordering can induce ferroelectricity}
\begin{figure}
\centerline{
\includegraphics[width=.5\columnwidth,angle=0]{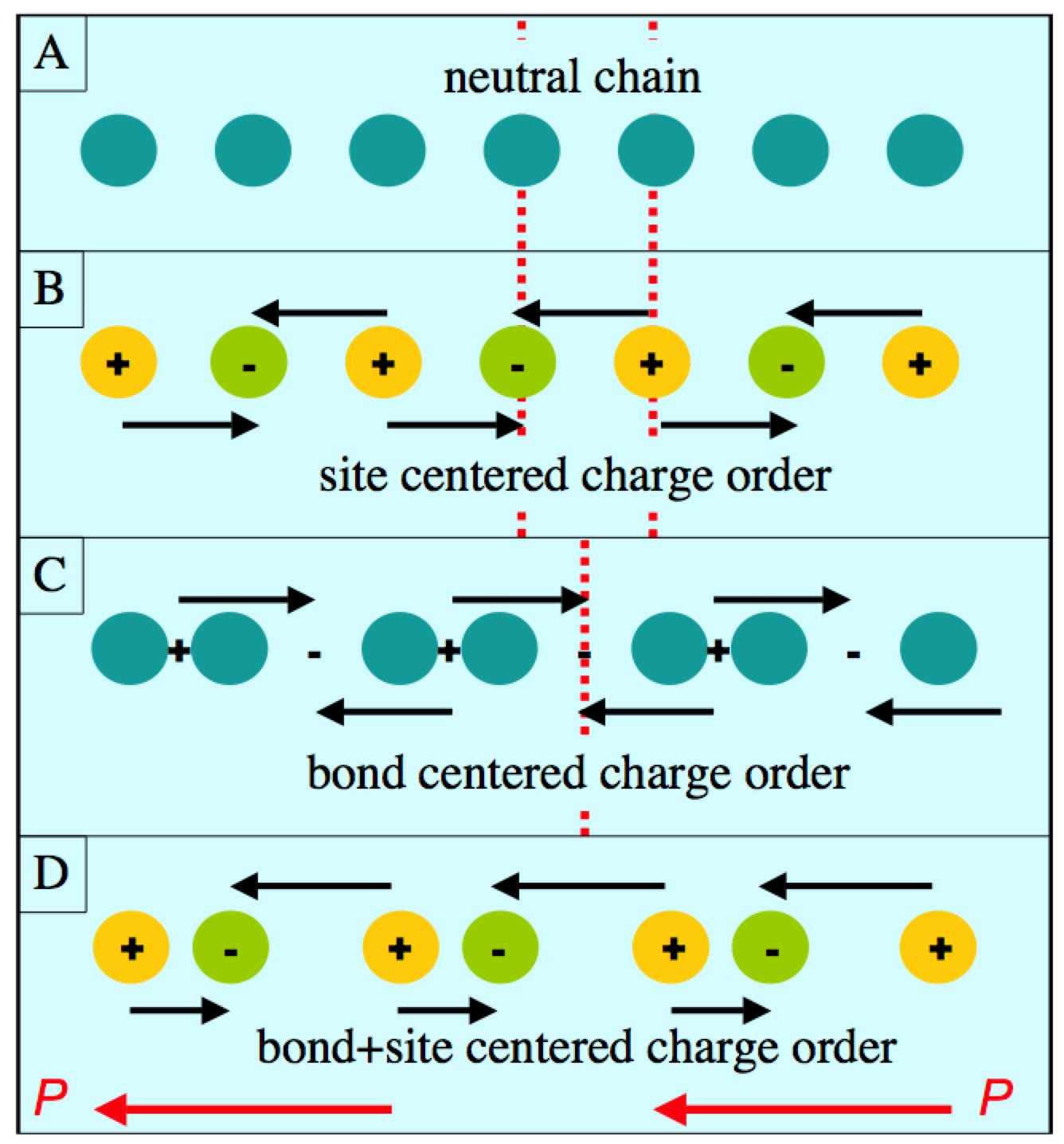}
}
\caption{(A) Example of a neutral one-dimensional chain that exhibiting (B) site-centered charge ordering, (C) bond-centered charge ordering, and (D) a linear combination of these two that is ferroelectric. The arrows indicate the polarization, which is in total zero in (B) and (C), but develops a macroscopic moment, indicated by the red arrow in (D). The red dashed lines in (A), (B) and (C) indicate mirror planes of the system.}
\label{fig:bond-site-cdw}
\end{figure}

The essential mechanism by which charge ordering can lead to the appearance of ferroelectricity is easily explained with the help of the schematic picture shown in Fig.~\ref{fig:bond-site-cdw}. In Fig.~\ref{fig:bond-site-cdw}A a homogeneous crystal (a one-dimensional chain in this case) with equal (say zero) charge on each site is shown. Fig.~\ref{fig:bond-site-cdw}B shows the same chain after a charge ordering in which sites become inequivalent, one set of sites has charge +e and the other -e, as in NaCl. This process does not breaks spatial inversion symmetry, so that the resulting state cannot have a net dipole moment. This is made explicit in Fig.~\ref{fig:bond-site-cdw}B by marking mirror planes of the charge ordered structure. 

Another type of charge ordering occurs when a system dimerizes, see Fig.~\ref{fig:bond-site-cdw}C. Such a lattice dimerization can have different origin, e.g. a Peierls distortion. In this case the sites remain equivalent, but {\it the bonds} are not, as the strong and weak bonds alternate. One can use the terminology of a site-centered charge ordering, or site-centered charge density wave (S-CDW) in the case of Fig.~\ref{fig:bond-site-cdw}B and a bond-centered CO, or bond-centered charge density wave (B-CDW) in the case of Fig.~\ref{fig:bond-site-cdw}C. Also the B-CDW structure is centrosymmetric and thus cannot be ferroelectric.

If one now combines both types of charge ordering in one system, the situation changes drastically. The situation with simultaneous site- and bond-centered CO is schematically shown in Fig.~\ref{fig:bond-site-cdw}D. Clearly inversion symmetry is broken in this case and each "molecule" (short bond in Fig.~\ref{fig:bond-site-cdw}D) develops a net dipole moment, so that as a result the whole system becomes FE. Thus solids can become ferroelectric if on top of site centered charge ordering also a bond dimerization occurs~\cite{efremov04}.

\subsection{Magnetic materials in which charge ordering can induce ferroelectricity}
We will see below that the simultaneous presence of inequivalent {\it sites} and {\it bonds}  can have a number of different origins. In some materials bonds are inequivalent just because of the crystallographic structure, and a spontaneous CO that occurs below a certain ordering temperature drives the inequivalence of the sites. Or {\it vice versa}, the material can contain ions with different valence, which after a structural dimerization transition induce FE. These two effects may also occur simultaneously, in which case there is one common phase transition.

The appearance of charge ordering is in itself quite ubiquitous in transition metal compounds. It is often observed in systems with ions that formally have a mixed valence. For instance half-doped manganites like La$_{1/2}$Ca$_{1/2}$MnO$_3$ or Pr$_{1/2}$Ca$_{1/2}$MnO$_3$ have one extra electron (or hole) per two Mn's, showing charge ordering of formally Mn$^{3+}$(d$^4$)  and Mn$^{4+}$(d$^3$). In these manganites CO typically extends over a large part of the doping phase diagram. For example in Pr$_{1-x}$Ca$_x$MnO$_3$ CO exists for $0.3 < x < 0.85$. For  $0.3< x<0.5$ the ordering is commensurate with the same periodicity as for $x=0.5$, but for $x>0.5$ it becomes incommensurate. Magnetic order sets in at temperatures below the CO temperature. The idea of charge ordering developing a ferroelectric component in such doped manganites was first put forward in Ref.~\cite{efremov04}.

Iron-oxides such as magnetite (Fe$_3$O$_4$), which exhibits the famous Verwey transition~\cite{verwey39,verwey41}, and LuFe$_2$O$_4$ \cite{ikeda00} have a charge ordering due to the presence of  both Fe$^{2+}$(d$^6$) and Fe$^{3+}$(d$^5$). In rare earth (R) nickelates of the type RNiO$_3$ the valence of nickel is 3+, but below a certain temperature a charge disproportionation into formally Ni$^{2+}$(d$^8$) and  Ni$^{4+}$(d$^6$) takes place~\cite{alonso99,mizokawa00}.

Traditionally one has site-centered ordering in mind when considering CO in such TM compounds. On the most basic level one can view this ordering as an alternation of TM ions with different valencies. However bond-centered CO is also a real possibility. For instance, in many quasi-one-dimensional compounds with partial electron occupation a Peierls transition occurs, which in fact can be viewed as a bond-centered CO, or B-CDW~\cite{brazovskii08}. Generally speaking, this possibility can be also realized in three-dimensional TM compounds like manganites or magnetite. There are however no general rules that tell us whether and when this situation occurs. Each time this question has to be studied separately on a microscopic level.

\begin{figure}
\centerline{
\includegraphics[width=.6\columnwidth,angle=0]{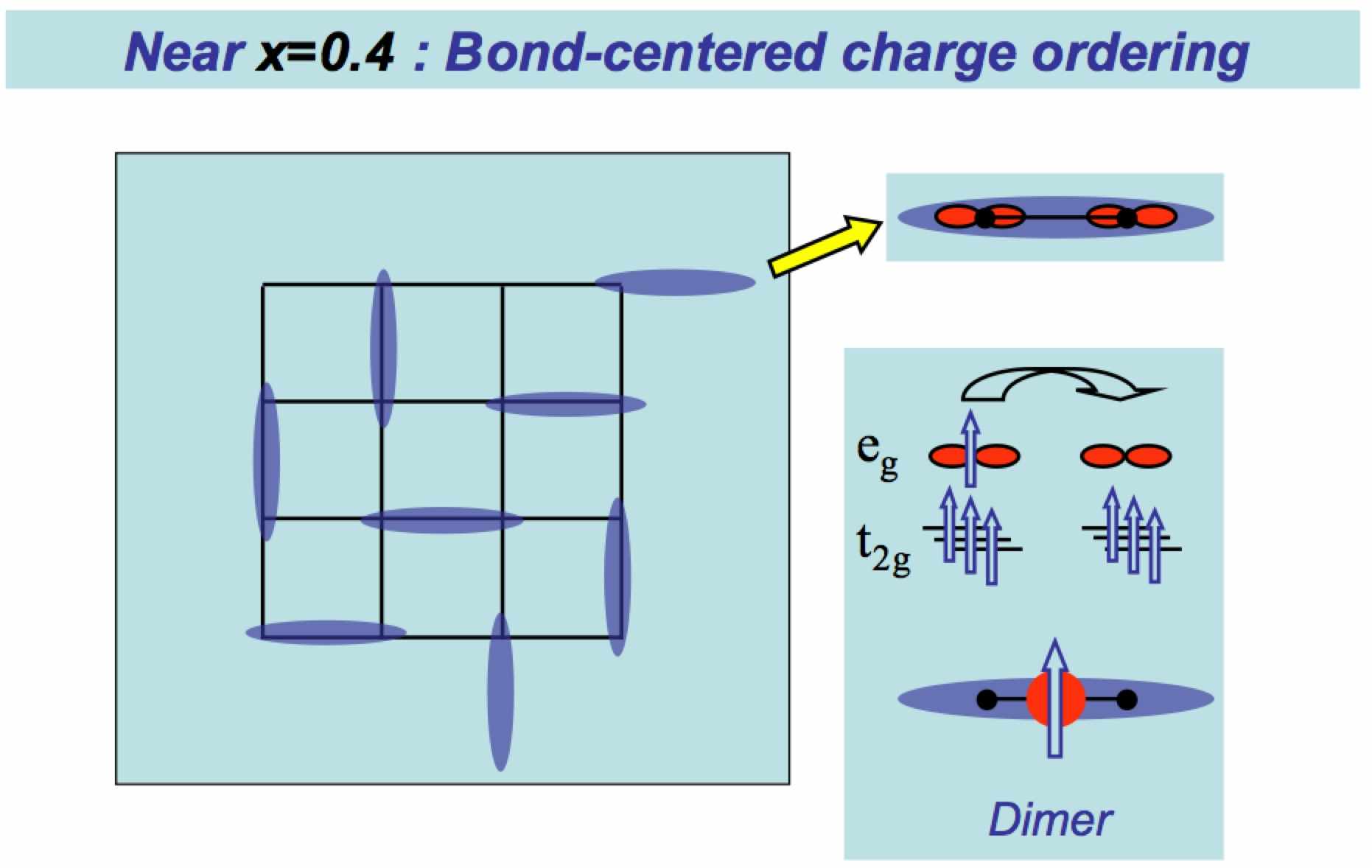}
\includegraphics[width=.6\columnwidth,angle=0]{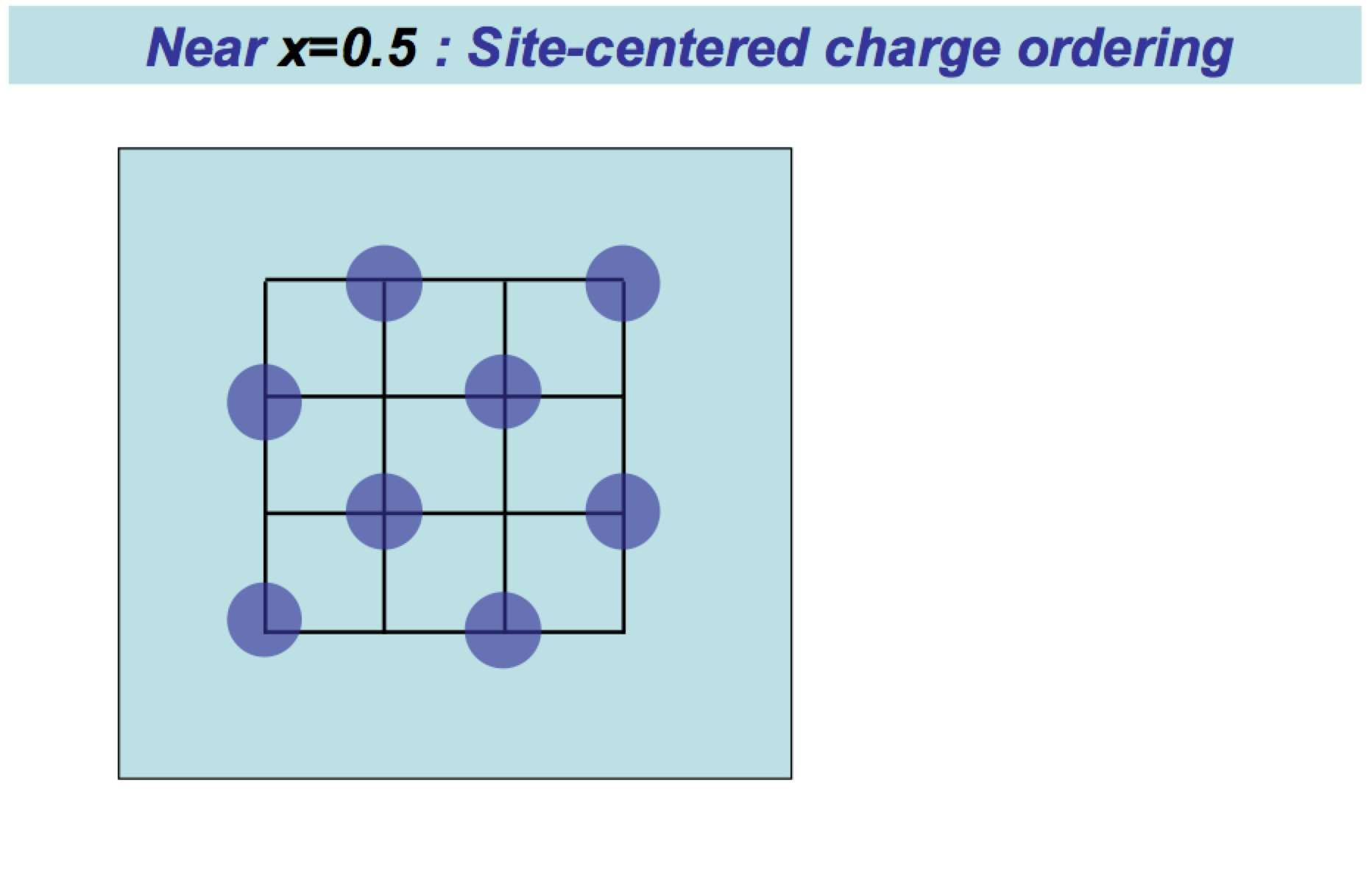}
}
\caption{Different types of charge ordering in doped manganites. Left: bond-centered charge ordering of "Zener polarons" proposed by Daoud-Aloudine et al.~\cite{daoud02}. Right: site-centered CO as proposed in the 1950's~\cite{wollan55,goodenough55}.}
\label{fig:manganite_co}
\end{figure}

\section{Type I Multiferroics with Ferroelectricity due to Charge Ordering}
\subsection{Perovskite Manganites of the type Pr$_{1-x}$Ca$_x$MnO$_3$} 
The question whether one or the other (i.e. site- or bond-centered) charge ordering structure (see Fig.~\ref{fig:manganite_co}) is realized in (PrCa)MnO$_3$ and whether a ferroelectric state can result from it is experimentally rather controversial. We will review this discussion after presenting the theoretical arguments and calculations that point into the direction of a possible multiferroic groundstate in these manganites.

\subsubsection{Theoretical situation}
\begin{figure}
\centerline{
\includegraphics[width=.4\columnwidth,angle=0]{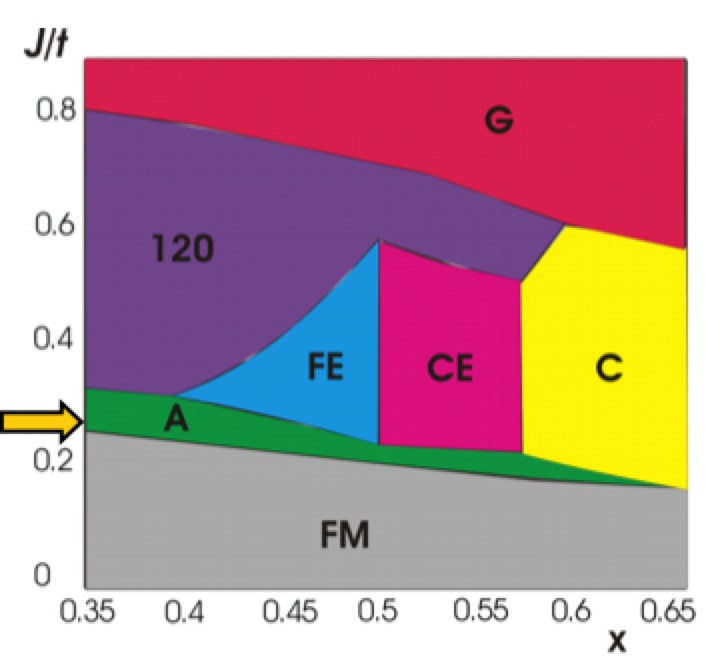}
\includegraphics[width=.5\columnwidth,angle=0]{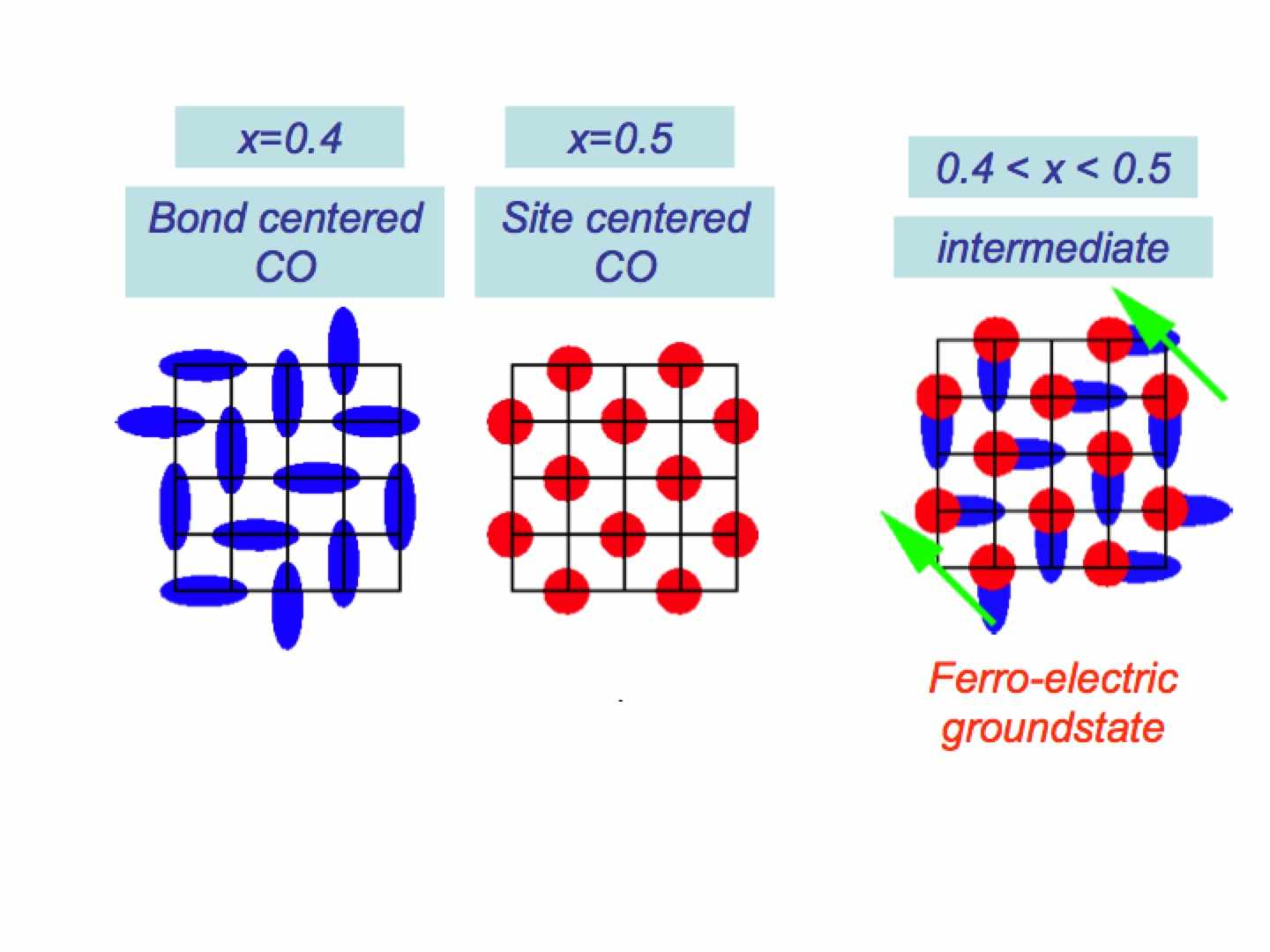}}
\caption{Left: Phase diagram of the degenerate double exchange model, by~\cite {efremov04}. 
The arrow indicates a value of the ratio of the effective superexchange $J$ and hopping $t$ for which the sequence of phases as a function of doping is the experimental one. The letters G, 120, A, C, CE, FM and FE denote, respectively, the magnetic G, 120 degree Jaffet-Kittel, magnetic A, C and CE phase, the ferromagnetic (FM) and ferroelectric (FE) phase. Right: schematic representation of the site- and bond-centered charge ordered states that result in an intermediate ferroelectric state if both types of charge ordering are present simultaneously.}
\label{fig:phasediagram}
\end{figure}

In Ref.~\cite{efremov04,efremov05_1} the question of site-CO versus bond-CO was treated theoretically within a double exchange model that takes into account the double degeneracy of active $e_g$-electrons, as well as the underlying magnetic structure. This approach had proved to be quite successful for overdoped and half-doped  manganites before~\cite{brink99_1,brink99_2,brink04}, and could later even be extended to undoped LaMnO$_3$~\cite{efremov05_2}, which are usually considered as Mott insulators. The results of the computations in the framework of the degenerate double exchange model are presented in the phase diagram of Fig.~\ref{fig:phasediagram}. The most important result here is the blue "triangle" to the left of $x \sim 0.5$: in the simplest approximation in this part of the phase diagram the bond-centered CO is lower in energy than the usual site-centered one. The observation that these different charge ordered states are in general very close in energy is confirmed independently by density functional calculations~\cite{patterson05}.

A more detailed analysis shows that the situation here is actually more complicated and much more interesting: in this region there occurs a {\it coexistence of site- and bond-centered CO}. According to our general picture, such a state is FE, see  right panel of Fig.~\ref{fig:phasediagram}. In this "triangular" part of the phase-diagram the character of the CO actually changes in a regular way: it evolves from a pure site-centered groundstate with magnetic CE-type order at $x=0.5$, to an admixture of bond-centered state increasing with decreasing $x$, and finally to a pure bond-centered one at the left edge of the "triangle" in Fig.~\ref{fig:phasediagram}.  One more point is worth mentioning. The theoretical treatment was done for groundstates with long-range magnetic ordering. Experimentally CO in Pr$_{0.6}$Ca$_{0.4}$MnO$_3$ sets in at $\sim$235 K, whereas the Ne\'el temperature is significantly lower, T$_N \sim 160$ K. However, a recent study indicates that magnetic correlations survive in this system much above the temperature of a long-range three-dimensional magnetic ordering -- actually up to T$_{CO}$~\cite{fernandez-baca}. Such short-range magnetic correlations are sufficient for our theoretical approach to be valid.

\subsubsection{Experimental situation}
For manganites the possibility of a bond-centered charge ordering was first proposed in Ref.~\cite{daoud02}, under the name of a "Zener polaron" state. On the basis of a detailed structural study of single crystalline Pr$_{1-x}$Ca$_x$MnO$_3$  (with $x \sim 0.4$) it was proposed that, contrary to the accepted charge ordering picture~\cite{wollan55,goodenough55}, shown in the center of the right panel of Fig.~\ref{fig:manganite_co}, the low-temperature phase of this system is better described not as a site-centered, but as a bond-centered CO. In this bond-centered CO state one extra electron is shared by the pair of neighboring Mn ions, moving back and forth between them and thus, by the double exchange mechanism, orienting the localized spins of their t$_{2g}$ electrons parallel. Hence the terminology "Zener polaron" -- a two-site polaron with ferromagnetic coupling. The proposed bond-centered CO structure of Ref.~\cite{daoud02} is schematically illustrated in the left panel of Fig.~\ref{fig:manganite_co}.

The question whether "Zener polarons" are present in these doped manganites is hotly debated in the literature. Whereas on the basis of a detailed single-crystal structural study the authors of Ref.~\cite{daoud02} claim that the bond-centered structure is realized at least for $x=0.4$, resonant elastic X-ray scattering experiments shown that for $x \sim 0.4 - 0.5$ Mn ions are inequivalent, which was interpreted by Grenier and coworkers as a disproof of the Zener polaron picture~\cite{grenier04}.  A recent high-resolution transmission electron microscopy and electron-diffraction study, on the other hand, confirms the presence of a "Zener polaron" state in Pr$_{1-x}$Ca$_x$MnO$_3$~\cite{wu07}.

From the theoretical considerations presented above it seems that this controversy is to an extent artificial, as it is based on an oversimplified treatment of the bond-centered CE structure. First of all, theoretical results show that at $x=0.5$ indeed the conventional site-centered structure with inequivalent Mn ions is realized. But even for $x=0.4$, where, according to our results~\cite{efremov04}, Mn-Mn bonds become inequivalent, simultaneously there exists also a site-centered CO, i.e. also the Mn ions should be inequivalent. Thus, the theoretical results reconcile the pictures of  Refs.~\cite{daoud02} and~\cite{grenier04} and show that the actual situation is in between, as it combines features of both bond-centered (Zener polaron) and site-centered (checkerboard type) CO. The fact that the actual symmetry of Pr$_{0.6}$Ca$_{0.4}$MnO$_3$ is non-centrocymmetric P11m~\cite{daoud02}, agrees with this picture and, most importantly, with the conclusions that this system should be multiferroic.

A direct observation of ferroelectricity in Pr$_{1-x}$Ca$_x$MnO$_3$ could settle the case. Unfortunately the build-up of a macroscopic polarization is hindered by the finite conductivity of this system, also precluding direct measurement of the polarization. There are at the moment two experimental studies in which an anomaly (a peak) in the dielectric constant $\epsilon$  was observed at the CO transition temperature~\cite{rivadulla, mercone}, strongly suggesting the presence of ferroelectricity.  Two other recent experiments were also interpreted as confirming the presence of FE in  Pr$_{1-x}$Ca$_x$MnO$_3$~\cite{jooss07,amaral}.

The anomaly of the dielectric constant at the CO transition was also found in a related system Pr$_{1-x}$Na$_x$MnO$_3$ for $x$=0.21, which correspond to a hole doping n$_h$=0.42~\cite{sanchez08}. The authors argue that it is a bulk effect, and interpret it as a possible indication of ferroelectricity in this system.

\subsection{Fe$_3$O$_4$: Multiferroic Magnetite}
The manganites that we discussed in the previous section are not the only, and possibly even not the best example of ferroelectric and multiferroic behavior that is caused by charge ordering. Another intriguing and famous case is magnetite, Fe$_3$O$_4$ -- the first magnetic material known to the mankind, see e.g.~\cite{mattis81}, with a ferrimagnetic ordering occurring already below $\sim$860 K. It is also famous because of what apparently is the first example of an insulator-metal transition in TM oxides --   the Verwey transition at T$_V$=120 K~\cite{verwey39,verwey41}, a transition that in spite of almost 70 years of dedicated research is still not completely understood. It is much less known, however that Fe$_3$O$_4$ is {\it ferroelectric} in the insulating state below the Verwey temperature T$_V$~\cite{rado75,rado77,miyamoto88,miyamoto93,miyamoto94}. So in addition to all its acclaimed fame, magnetite is probably also one of the first multiferroics.

\subsubsection{Charge ordering in magnetite}
\begin{figure}
\centerline{
\includegraphics[width=.45\columnwidth,angle=0]{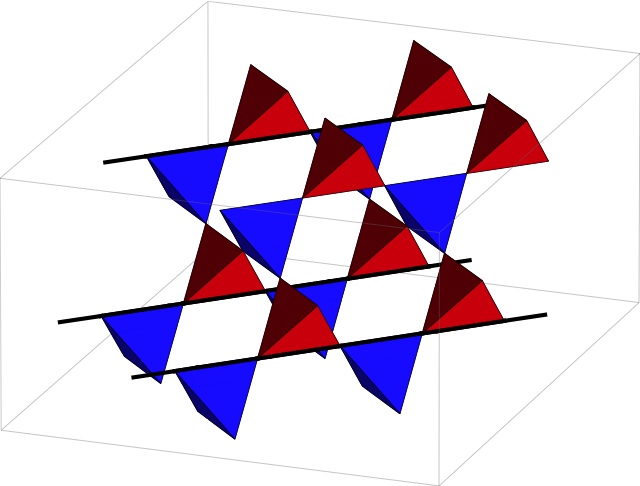}
\includegraphics[width=.45\columnwidth,angle=0]{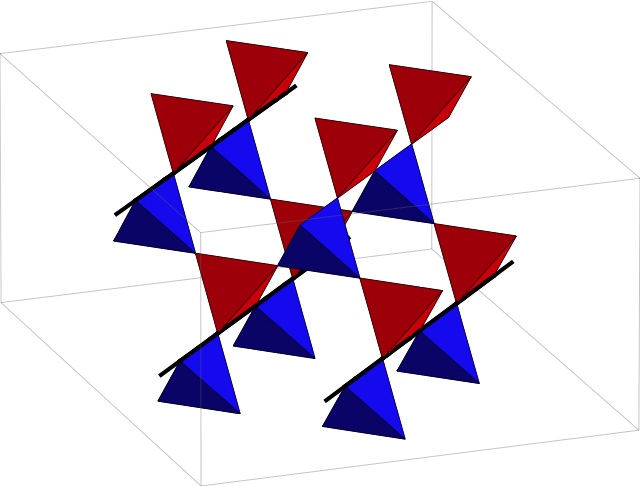}}
\caption{Pyrochlore lattice made by the B-sites of a spinel structure from two different viewpoints. In Fe$_3$O$_4$ equal numbers of formally Fe$^{3+}$ and Fe$^{2+}$ ions occupy these sites.}
\label{fig:pyrochlore}
\end{figure}

Fe$_3$O$_4$ crystallizes in an inverse cubic spinel structure with two distinct iron positions. The iron B sites are inside an oxygen octahedron and contain 2/3 of the Fe ions, with equal numbers of Fe$^{3+}$ and Fe$^{2+}$. These sites by themselves form a pyrochlore lattice, consisting of a network of corner-sharing tetrahedra, see Fig.~\ref{fig:pyrochlore}. The iron A sites contain the other one third of the Fe ions and are not considered relevant for the charge ordering physics. The Verwey metal-insulator transition at T$_V$=120K is apparently related to charge ordering transition of the two types of charges that live on the sites of the pyrochlore lattice of the iron B sites. The originally proposed charge ordering pattern~\cite{verwey41} consists of an alternation of Fe$^{2+}$ and Fe$^{3+}$ ions in the $xy$-planes at the B-sites and was later shown to be too simple. Numerous much more complicated CO patterns were proposed, see e.g. Ref.~\cite{phil_mag80} and references therein. The detailed charge ordering pattern is still unresolved. 

The difficulty in determining the charge ordering structure is related to the strong frustration of simple bipartite ordering on a pyrochlore lattice, first discussed by Anderson~\cite{anderson56}. This frustration has the consequence that there is a macroscopic number of charge configurations with the same groundstate energy, if only nearest neighbor Coulomb interactions are taken into account. This huge degeneracy  remains even in the presence of additional constraints on the possible charge ordering patterns, the best known of these is the so-called Anderson constraint, where each tetrahedron is required to have two  Fe$^{2+}$ and two Fe$^{3+}$ ions. Such a macroscopic degeneracy is expected to contribute significantly to the entropy of the system. It was noticed very early on that this situation is very similar to the frustrated order in (water) ice, in which the hydrogen atoms in ice are arranged and constrained according to the so-called Bernal-Fowler rules~\cite{pauling33,bernal33}. Also in the problem of spin-ice frustration is a dominant factor, and its consequences have recently attracted much attention, see for instance~\cite{castelnovo08}.

It is clear that residual interactions, beyond the nearest neighbor Coulomb repulsion, have to be present. These are important and relevant because below T$_V$ magnetite is ordered into a unique groundstate. Moreover, a recent high resolution neutron and X-ray diffraction study by Wright and coworkers~\cite{wright01} concludes that the Anderson rule is violated in Fe$_3$O$_4$. But our main point here is that even in the very careful structural investigation of Wright and coworkers a centrosymmetric monoclinic group is used to analyze the diffraction data, which precludes the possibility of a ferroelectric groundstate. Thus the real structure of Fe$_3$O$_4$ is apparently even more complicated than the ones discussed so far. 

\begin{figure}
\centerline{
\includegraphics[width=.6\columnwidth,angle=0]{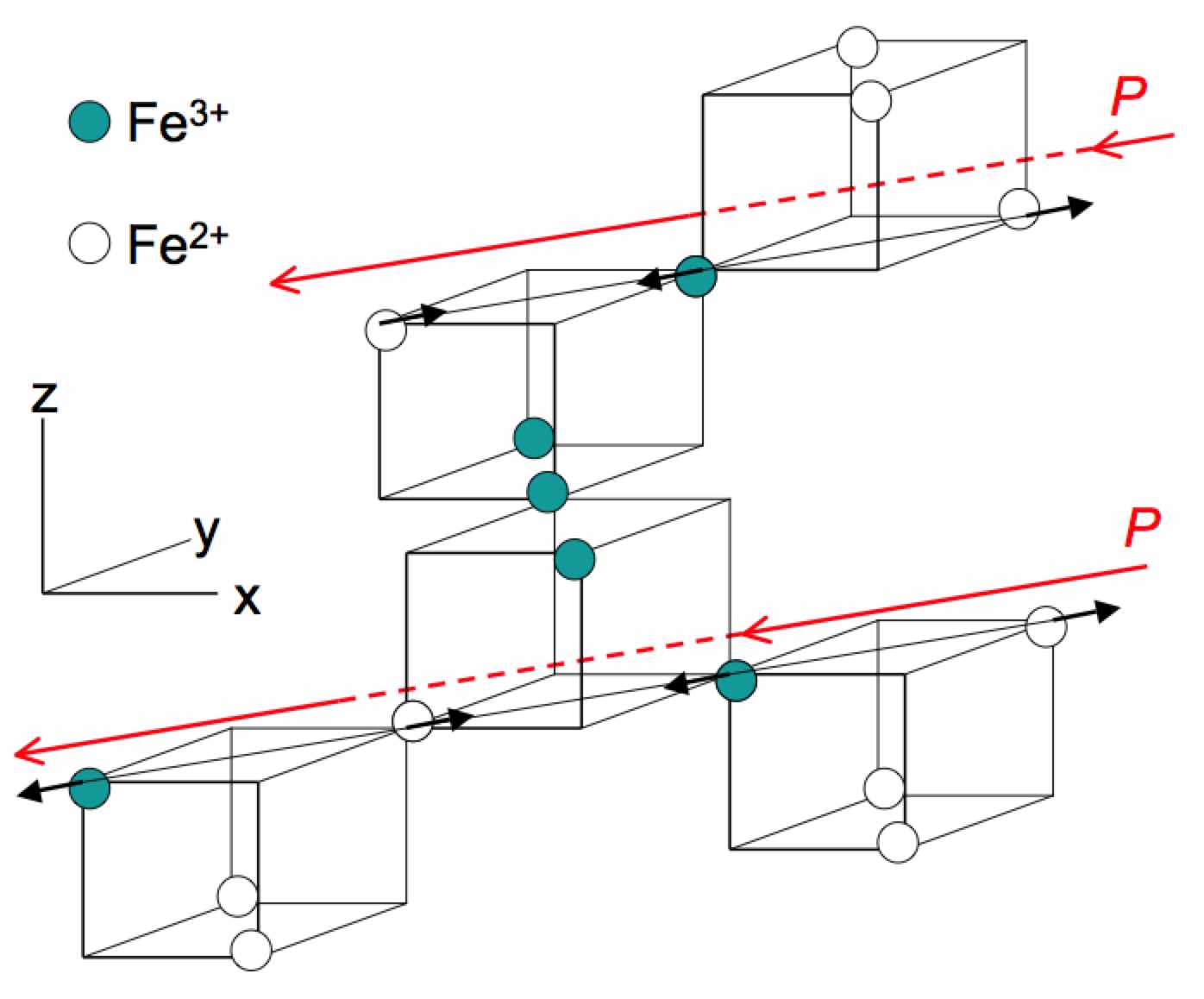}
}
\caption{Illustration of the possible origin of ferroelectricity in magnetite. Emphasized are the Fe chains running in [110] direction of magnetite -- a pyrochlore lattice made by the spinel B-sites. In the $xy$ chains there is an alternation of Fe$^{2+}$ and Fe$^{3+}$  ions (open and filled circles). Simultaneously there is an alternation of short and long Fe-Fe bonds; shifts of Fe ions are shown by black arrows. The resulting polarization is indicated by the red arrows.}
\label{fig:Fe3O4}
\end{figure}

\subsubsection{Possible origin of ferroelectricity in magnetite}
According to Ref.~\cite{miyamoto93}, the electric polarization $P$ of Fe$_3$O$_4$ below the Verwey transition lies in the $b$-direction. The polarization can be switched by electric field and it leads to the formation of a ferroelectric domain structure that can only be explained by assuming a triclinic crystal structure~\cite{medrano}. 

The microscopic origin of ferroelectricity in magnetite remains to be unveiled, but one can argue~\cite{khomskii06, khomskiiAPS} that most probably it is related to the coexistence of site-centered and bond-centered CO of a type that was discussed for the manganites above. Actually a detailed analysis of  the structural data of Ref.~\cite{wright01} shows that the situation in this system may be not so far from the one illustrated in the right panel of Fig.~\ref{fig:phasediagram}. Indeed, in the proposed structural and charge pattern one notices that, besides the site-centered CO -- the alternation of the formal Fe$^{2+}$ and Fe$^{3+}$ valence states --  there is also a strong modulation of the Fe-Fe distances (the bond lengths). In this structure, shown in Fig.~\ref{fig:Fe3O4}, one sees that e.g. along the Fe chains running in the [110] direction (in cubic setting), i.e. in the $xy$ chains, there is an alternation of Fe$^{2+}$ and Fe$^{3+}$  ions (open and filled circles in Fig.~\ref{fig:Fe3O4}), but simultaneously there is an alternation of short and long Fe-Fe bonds. This direction corresponds to the monoclinic $b$-direction, in which the polarization is observed. In this framework each of such [110] mixed bond- and site-centered CO chains gives a non-zero contribution to the electrical polarization, cf. Fig.~\ref{fig:phasediagram}.

In the analysis of the structural data in Ref.~\cite{wright01} the enforced centrosymmetry is accommodated by a certain pattern of [110] mixed chains that changes into the opposite pattern in the "upper part" of the unit cell that is doubled in the $c$-direction. In the enforced centrosymmetric structure Fe$_3$O$_4$ is therefore antiferroelectric. How to change the pattern of Ref.~\cite{wright01} to incorporate ferroelectricity is not entirely clear at present, but it seems that the main structural features point very strongly in the direction of the mechanism proposed in Ref.~\cite{efremov04} -- the coexistence of bond-centered and charge-centered charge ordering -- as the mechanism for ferroelectricity in magnetite. 

\subsection{Frustrated and Charge Ordered LuFe$_2$O$_4$}
Recently multiferroicity was observed in LuFe$_2$O$_4$~\cite{ikeda00,ikeda05}. Despite the similarity of its chemical formula to the one of a spinel, the structure of LuFe$_2$O$_4$ is very different from a spinel.  It has a double layer structure, with a triangular iron lattice within each layer.  The average valence of Fe in it is Fe$^{2.5+}$, similar to the B sites in Fe$_3$O$_4$, that also nominally have an equal number of Fe$^{2+}$ and Fe$^{3+}$ ions, or one extra electron (or hole) per two sites. Accordingly, charge ordering may occur in LuFe$_2$O$_4$  and indeed  does occur at $T_c$=330 K.

\begin{figure}
\centerline{
\includegraphics[width=0.65\columnwidth,angle=0]{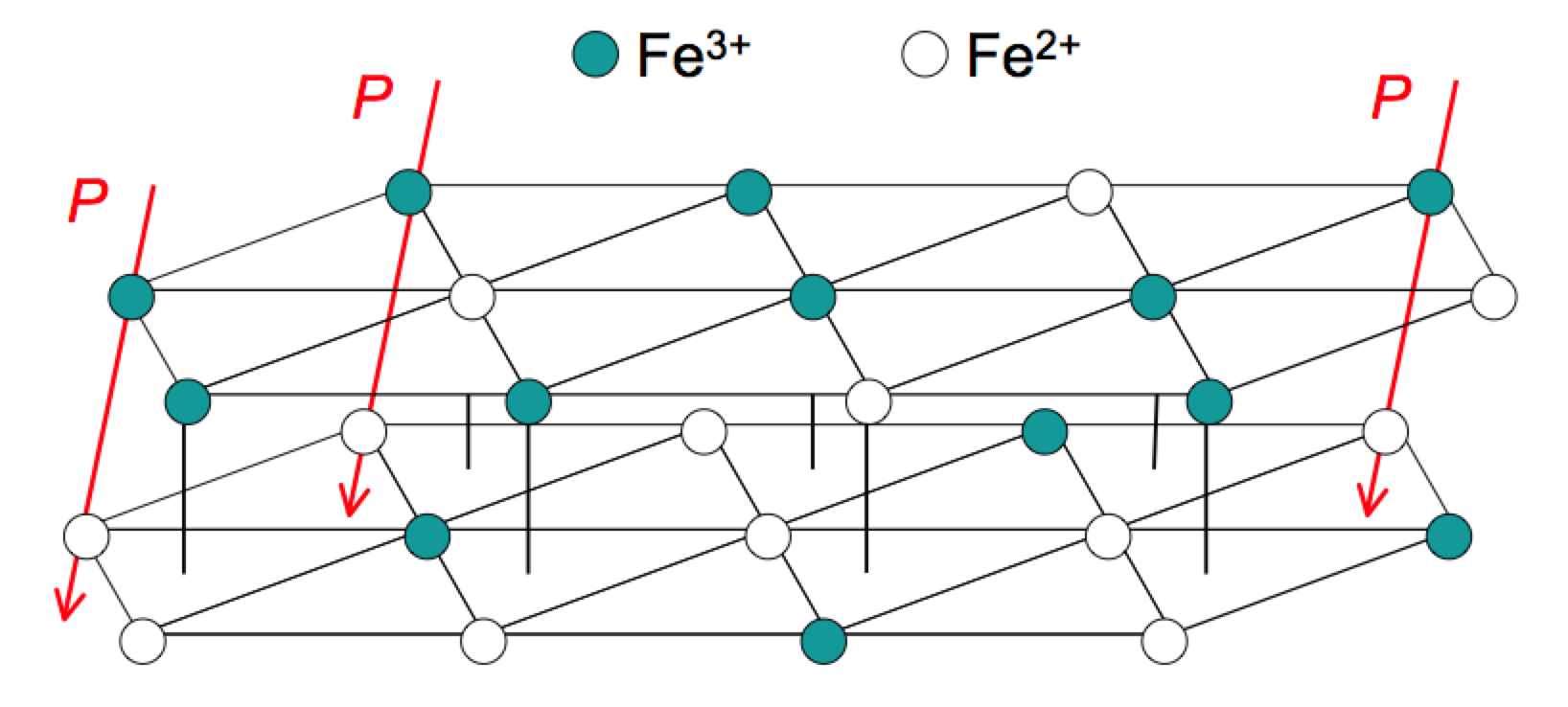}
}
\caption{Bilayer of the FeO$_2$ triangular lattices in LuFe$_2$O$_4$ with a schematic view of charge redistribution between the layers and the interlayer charge ordering that results in a macroscopic electric polarization indicated by the red arrows.} 
\label{fig:LuFe2O4}
\end{figure}

However, because of the frustrated nature of the triangular iron lattice in each FeO$_2$ layer, the usual bipartite checkerboard Fe$^{2+}$/Fe$^{3+}$ charge ordering is not favorable and another option becomes preferred. A charge redistribution between layers takes place, so that in each bilayer one layer, say the lower layer, has a 2:1 ratio of Fe$^{2+}$/Fe$^{3+}$  and its upper counterpart has the inverse 1:2 ratio. Because of this interlayer charge redistribution each triangular layer can have a perfect, unfrustrated charge ordering with three sublattices in each triangular layer, one sublattice being occupied e.g. by Fe$^{3+}$,  and two others (forming a honeycomb lattice with these Fe$^{3+}$ at the center of each hexagon) by Fe$^{2+}$. The charge ordering is {\it vice versa} in the other layer, see Fig.~\ref{fig:LuFe2O4}. As a result each {\it bilayer} acquires a dipole moment, shown in Fig.~\ref{fig:LuFe2O4}, and the total system becomes ferroelectric~\cite{ikeda00, ikeda05, nagano07}. As ferrimagnetic ordering sets in at T$_c$=250 K, LuFe$_2$O$_4$ is multiferroic below this temperature.

Thus the FE in LuFe$_2$O$_4$  is due to a combination of two factors: the bilayer character of the crystal structure, and the frustrated charge ordering leading to the formation of charged planes (e.g., negative charging of the lower layer and positive charging of the upper one). As, generally speaking, in such a situation CO may be very strong, one could in principle expect large spontaneous polarizations -- much larger than for example in type-II multiferroics with magnetically-driven ferroelectricity~\cite{mostovoy07}. And indeed, the value of the polarization in LuFe2O4 is 0.24 $C/m^2$ -- comparable to the polarization of 0.26 $C/m^2$ in BaTiO$_3$~\cite{jona62}.

\subsection{Quasi one-dimensional organic materials}
\begin{figure}
\centerline{
\includegraphics[width=.6\columnwidth,angle=0]{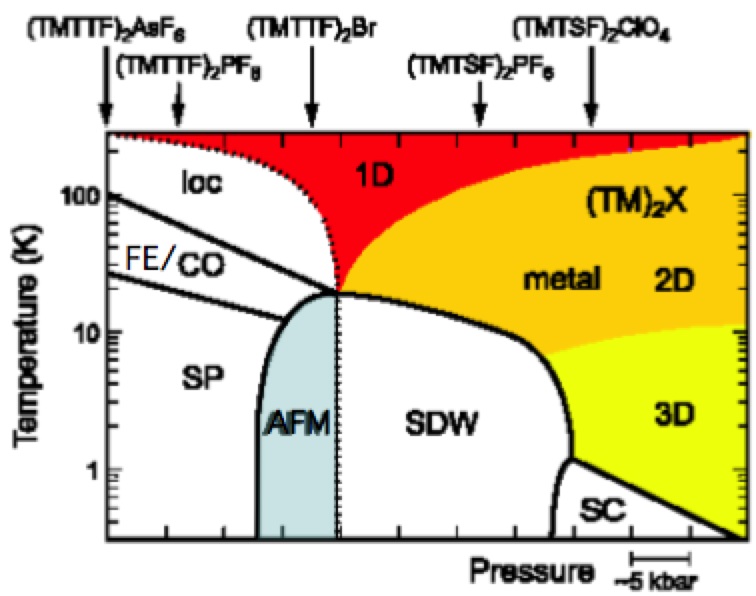}
}
\caption{Generic phase diagram of (TMTTF)$_2$X or (TMTSF)$_2$X molecular crystals, from Ref.~\cite{brazovskii08}. 
CO -- charge-ordered phase; SP -- spin-Peierls one; AFM -- the phase with antiferromagnetic ordering; SC -- superconducting phase. The charge-ordered phase is simultaneously ferroelectric (FE)~\cite{brazovskii}. In the blue shaded region multiferroicity is expected to occur.} 
\label{fig:phase1d}
\end{figure}

A situation that is similar to the one discussed in the previous section -- bonds inequivalent just because of the underlying crystal structure, and ferroelectricity caused by charge ordering -- exists also in other materials. Probably one of the first such examples is the quasi-one-dimensional organic system (TMTTF)$_{2}$X \cite{brazovskii}. In this material there is one electron (or rather one hole) per two sites (i.e. per two TMTTF molecules), but the molecular chains are dimerized due to a particular ordering of counter-ions X (X=PF$_6$, AsF$_6$, SbF$_6$), which gives rise to inequivalent bonds.  With decreasing temperature at T$_{CO} \sim $ 50-150 K charge ordering occurs, after which the molecules (sites) become inequivalent, having alternating charge. The situation then becomes exactly the same as the one shown in Fig.~\ref{fig:bond-site-cdw}D, and consequently the low-temperature phase of these materials becomes ferroelectric. This is clear from anomalies in the dielectric constant $\epsilon$, which above T$_c$ shows Curie behavior. Below T$_{CO}$ in the simplest picture electrons (or holes) are localized at every second molecule, and the corresponding localized spins either undergo a spin-Peierls transition or have long-range antiferromagnetic ordering when temperature decreases. This behavior is seen in the generic phase diagram of Fig.~\ref{fig:phase1d}, adopted from Ref.~\cite{brazovskii08}. The spin-Peierls transition would effectively remove magnetism from the picture, but the systems with antiferromagnetic ordering should be classified as multiferroic.
 
\section{Type-II Multiferroics with Ferroelectricity due to Charge Ordering and Magnetostriction}
Magnetically-driven FE in type-II multiferroics may have two microscopic origins. One of them, probably the most common, works in systems like TbMnO$_3$~\cite{kimura03}, Ni$_3$V$_2$O$_8$~\cite{lawes04}, MnWO$_4$~\cite{taniguchi06,heyer06,arkenbout06}, in multiferroic pyroxenes~\cite{jodlauk07} and in some other systems. The mechanism of FE in them is usually an inverse Dzyaloshinskii-Moriya effect~\cite{sergienko06}, which operates in systems with non-collinear, usually spiral magnetic structures of certain type~\cite{nagaosa,mostovoy07}. It requires the direct action of the relativistic spin-orbit interaction. 

A second possible mechanism works also for collinear magnetic structures and does not require the presence of spin-orbit coupling: it is based on magnetostriction. For the magnetosctriction to give multiferroic behavior one usually requires the presence of inequivalent magnetic ions, with different charges. These, in their turn, may be either just different TM ions, or the same element in different valence states. This mechanism for multiferroicity is illustration by Fig.~\ref{fig:Ising}. The situation here is almost identical with that shown in Fig.~\ref{fig:bond-site-cdw}B. Without spin ordering, the crystal structure is centrosymmetric. The magnetic structure itself is also inversion symmetric, but with a different inversion center.  Taking both charge and magnetic structures together, however, the system loses inversion symmetry. i.e. it can become FE.  The mechanism of creation of polarization would be the magnetostriction (MS). The MS is definitely different for ferro and antiferro bonds. If energy is gained by shortening of, for example, a ferromagnetic bond, the system will distort as shown in Fig.~\ref{fig:Ising}B, and one ends up with exactly the same situation as shown in Fig.~\ref{fig:bond-site-cdw}D, with unequal bonds and different charges at opposite ends of short "molecules". A simple model that exhibits this behavior is the Ising model with both nearest-neighbor and next nearest-neighbor interactions:

\begin{figure}
\centerline{
\includegraphics[width=.55\columnwidth,angle=0]{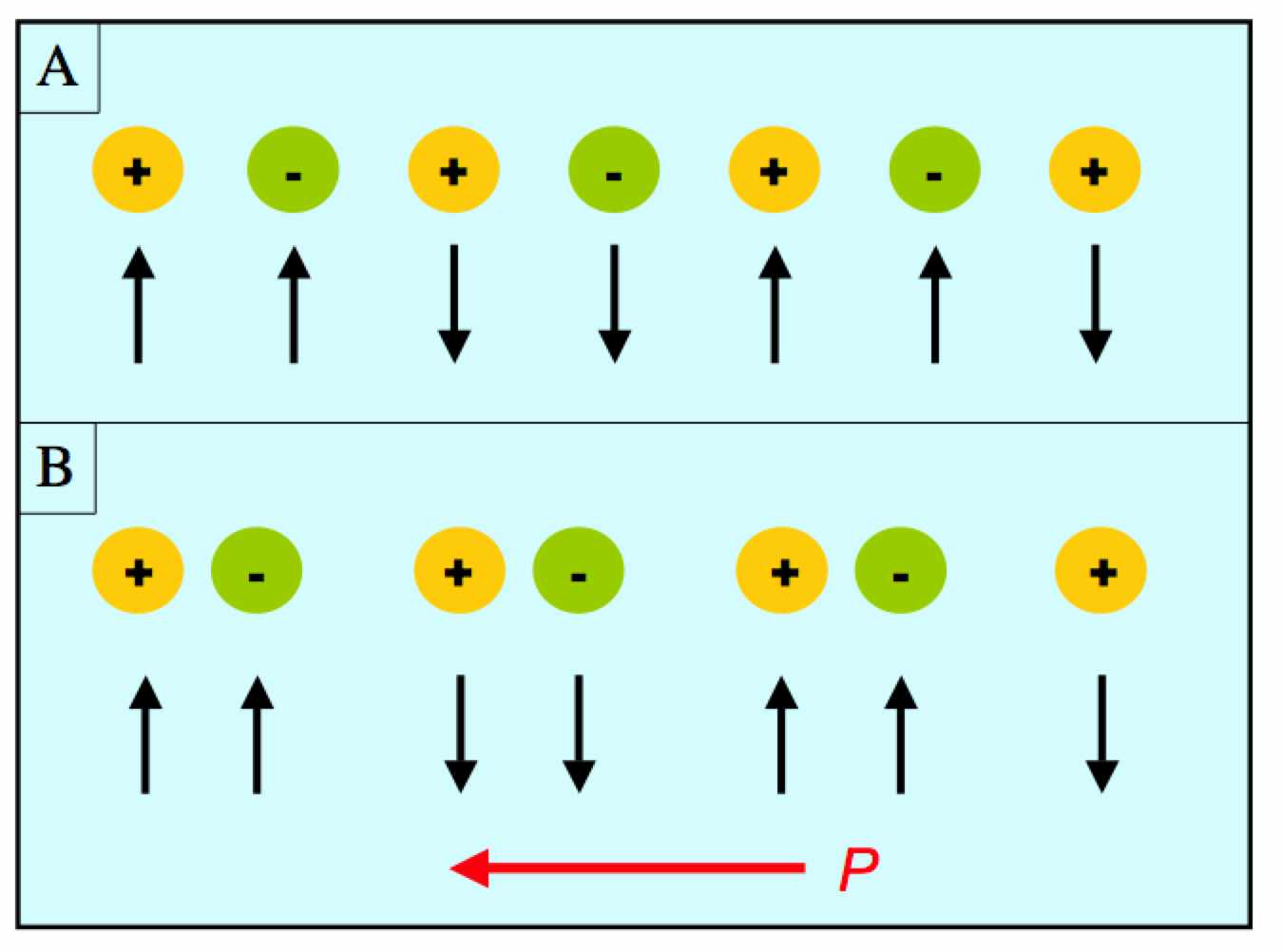}
}
\vspace{0cm}
\caption{(A) Chain with alternating charges and with the spin structure up/up/down/down. (B) effect of magnetostriction, shortening (in this example) the ferromagnetic bonds, producing a ferroelectric polarization, indicated by the red arrow in Fig.~\ref{fig:bond-site-cdw}D.}
\label{fig:Ising}
\end{figure}

\begin{eqnarray}
H = J_1 \sum_i S^z_iS^z_{i+1}  + J_2 \sum_i S^z_iS^z_{i+2}.
\end{eqnarray}

If the interaction $J_2$ is antiferromagnetic ($>0$) and sufficiently large, $J_2>1/2|J_1|$, the magnetic ordering will be of the up-up-down-down type, see Fig.~\ref{fig:Ising}. Depending on the sign of $J_1$, the bond with parallel spins (for $J_1<0$) or with antiparallel ones (for $J_1>0$) will shorten, increasing the magnetic energy gain on the corresponding bond. In both cases this will lead to a multiferroic state. An almost ideal realization of this scenario seems to be found recently in Ca$_3$CoMnO$_6$~\cite{choi08}. In this system with a quasi one-dimensional structure the ions Co$^{2+}$ and Mn$^{4+}$ alternate along the chain, and magnetic structure is of up-up-down-down type. And indeed, below $T_N = 16$ K, Choi {\it et al.} observe an electric polarization in this system~\cite{choi08}.

\subsection{Manganites of the type RMn$_2$O$_5$}
The manganites RMn$_2$O$_5$, where R is a rare earth ion~\cite{hur04}, are the first example of type-II multiferroics that were discovered. In these manganites ferroelectricity is presumably driven by magnetostriction. It can occur because the material contains Mn sites that are inequivalent by virtue of the crystallographic structure of the material, with on top of this a magnetically driven spontaneous distortion that makes bonds unequal and breaks inversion symmetry. RMn$_2$O$_5$  has a rather complicated crystal structure containing inequivalent Mn$^{3+}$ and Mn$^{4+}$ ions in different lattice positions with different oxygen coordination, with a magnetic ordering that breaks the inversion symmetry of the undistorted, non-ferroelectric lattice~\cite{betouras07,harris07}.

\begin{figure}
\centerline{
\includegraphics[width=.6\columnwidth,angle=0]{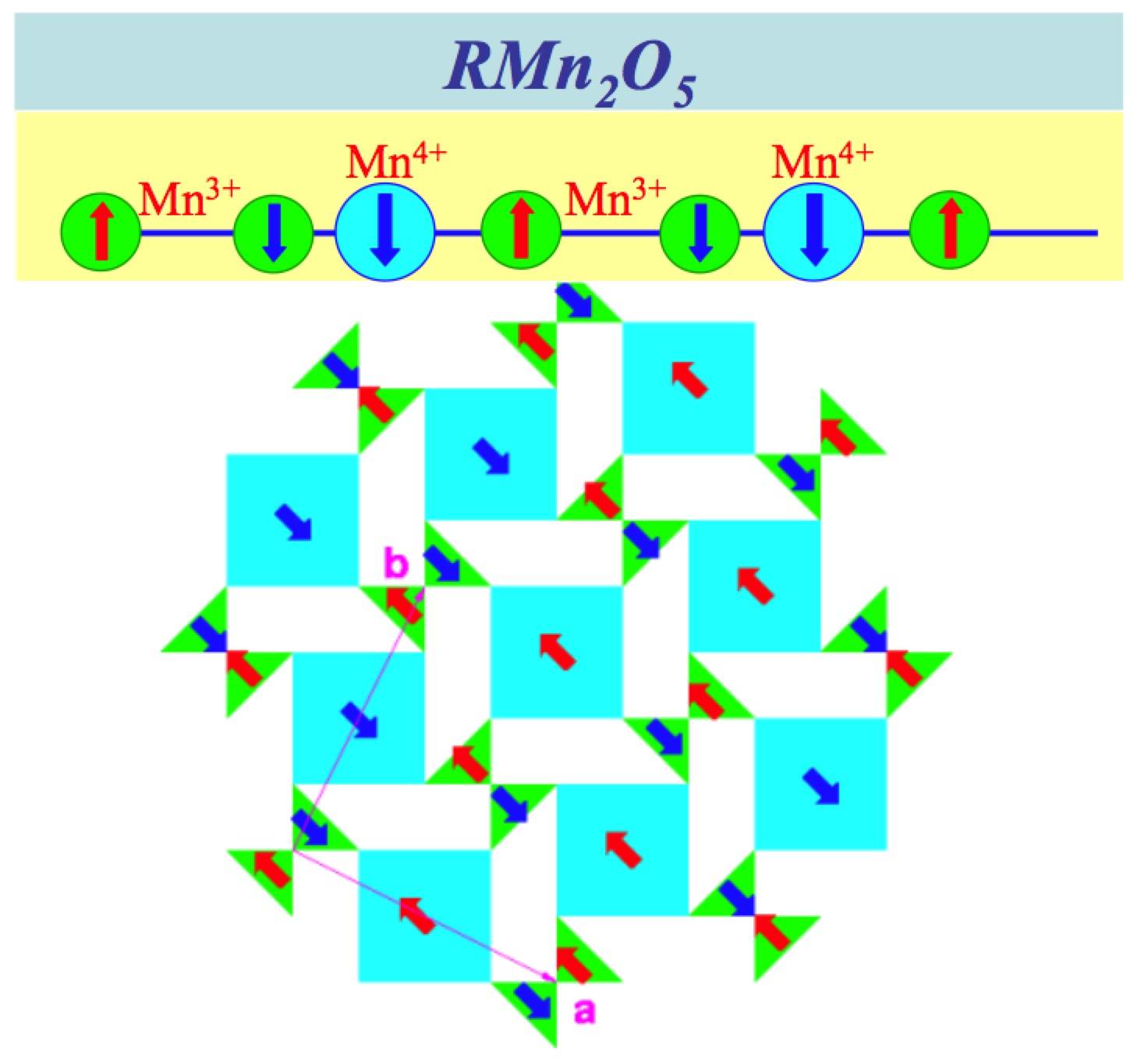}
}
\caption{Schematic view of the crystal structure of $R$Mn$_2$O$_5$ consisting of connected Mn$^{4+}$O$_6$ octahedra and Mn$^{3+}$O$_5$ pyramids; the figure is from the review of A.~B. Sushkov {\it et al} in  this issue. The chain of Mn$^{3+}$-Mn$^{3+}$-Mn$^{4+}$ along $b$-direction, with corresponding spin ordering, is shown in the upper panel.
}
\label{fig:HoMn2O5}
\end{figure}

Neutron and X-ray diffraction studies show that these manganites have space group \emph{Pbam} but it is expected that in their multiferroic state the actual symmetry group is {\emph{Pb}2$_1$\emph{m}}, which allows for a macroscopic electric polarization along the $b$ axis~\cite{Kagomiya,Chapon04,Alfonso}. The orthorhombic $Pbam$ crystal structure of HoMn$_2$O$_5$ consists of connected Mn$^{4+}$O$_6$ octahedra and Mn$^{3+}$O$_5$ pyramids (see Fig.~\ref{fig:HoMn2O5}). The octahedra share edges and form ribbons parallel to the $c$ axis. Adjacent ribbons are linked by pairs of corner-sharing pyramids. Below 38 K e.g. in HoMn$_2$O$_5$  a commensurate magnetic structure develops with propagation vector ${\bf k}=({1 \over 2},0,{1 \over 4})$, and simultaneously the system becomes ferroelectric~\cite{Blake}. 

Along the $b$-direction HoMn$_2$O$_5$ exhibits a charge and spin ordering that can schematically be denoted as a chain of Mn$^{3+}_{\Uparrow}$-Mn$^{4+}_{\Uparrow}$-Mn$^{3+}_\Downarrow$.  In the undistorted \emph{Pbam} structure the distances d$_{\Uparrow\Uparrow}$ (between Mn$^{3+}_\Uparrow$ and Mn$^{4+}_\Uparrow$) and d$_{\Uparrow\Downarrow}$ (between Mn$^{3+}_\Uparrow$ and Mn$^{4+}_\Downarrow$) are the same. Ab initio calculations of the relaxed structure~\cite{wang07,wang07_2,giovannetti07} reveal a shortening of distances between parallel spins Mn$^{3+}_{\Uparrow}$ and Mn$^{4+}_{\Uparrow}$ ions   -- in the ferroelectric {\emph{Pb}2$_1$\emph{m}} structure $d_{\Uparrow\Uparrow} < d_{\Downarrow\Uparrow}$, which optimizes the double exchange energy~\cite{mostovoy07,efremov04,betouras07}. 

\subsection{Origin of ferroelectricity in RMn$_2$O$_5$}
There is still some controversy about the microscopic mechanism of ferroelectricity in this system~\cite{mostovoy07}.  The most plausible explanation is the one indicated above: magnetically induced changes in bond lengths between Mn ions with different formal valence, adding up to a net ferroelectric moment. The actual picture that arises from microscopic electronic structure calculations supports this scenario~\cite{wang07,wang07_2,giovannetti07}, but is actually even more interesting, see below. An alternative picture is that the ferroelectricity in RMn$_2$O$_5$ is due to a spiral magnetic structure, like in many other type-II multiferroics~\cite{kimura07}, but most probably the weak spiral observed in~\cite{kimura07} is not the source, but rather the consequence of ferroelectricity, similar to the case of BiFeO$_3$.

\begin{figure}
\centerline{
\includegraphics[width=0.7\columnwidth,angle=0]{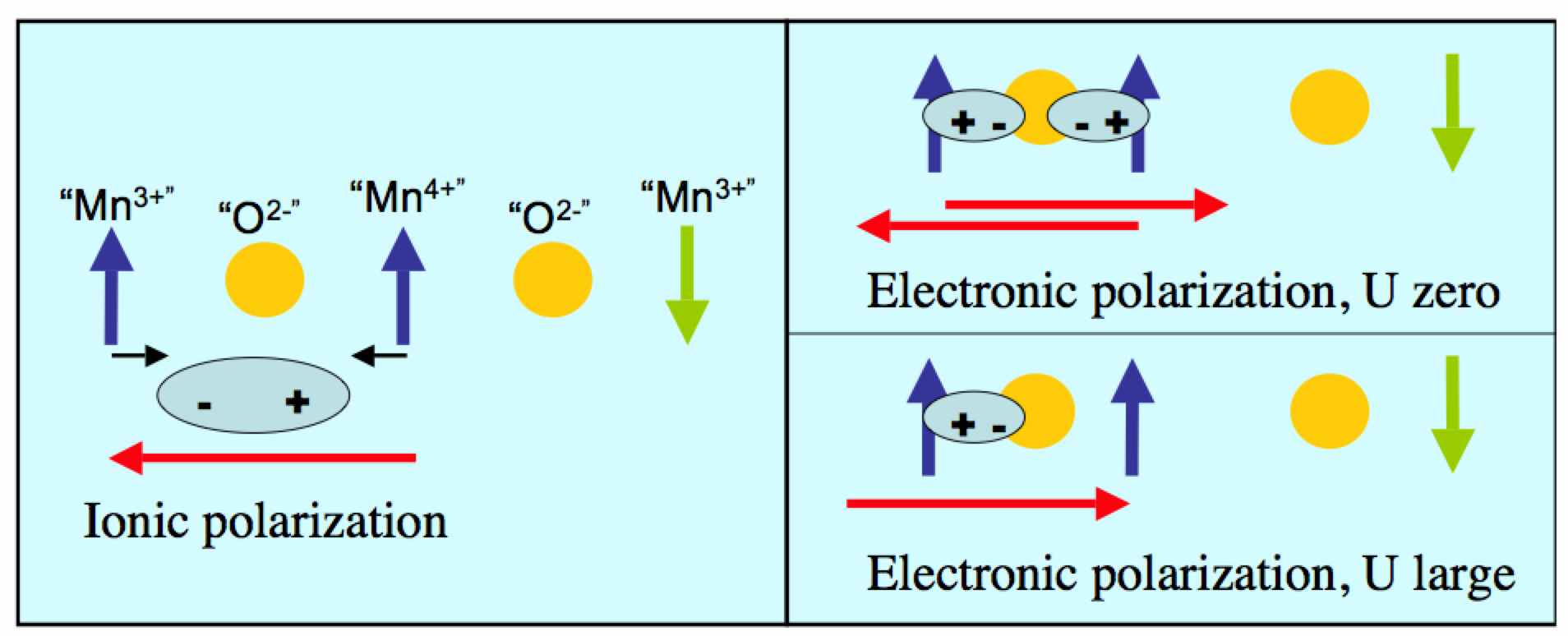}
}
\caption{Schematic view of the two contributions to the ferroelectric polarization in RMn$_2$O$_5$ in the uncorrelated ($U$=0) and strongly correlated limit (large $U$). In the latter the electronic polarization nearly cancels the ionic polarization. The labels "Mn$^{4+/3+}$" indicate Mn ions that have a valence of more/less than $3.5+$, respectively.
}
\label{fig:125polarization}
\end{figure}

The electronic structure calculations show that the magnetic ordering gives rise to a large {\it electronic} polarization, inducing a transfer of charge from the Mn-sites to the Mn-O-Mn bonds, similar in spirit to the perovskite manganites that we discussed in a previous section. This charge transfer drives the bond-length shortening and gives rise to an ionic polarization as the two Mn ions involved have formally different charges. The transferred charge itself, however, does not induce a large dipole moment, because the charges accumulated on the bonds comes from opposite directions in roughly the same amount~\cite{wang07,wang07_2,giovannetti07}. However, this changes drastically when the on-site Coulomb interactions between the Mn 3$d$ electrons are accounted for~\cite{giovannetti07}.  It depletes electrons from the Mn$^{4+}$ site, making it almost closed shell $t_{2g}$ and as a consequence its polarization cloud disappears. The overall result is that in these strongly correlated multiferroic manganites a substantial {\it electronic} polarization develops, which is almost as large as the {\it ionic} polarization, but opposite in direction, see Fig.~\ref{fig:125polarization}. In the end a small net polarization of  $P$=82 nC/cm$^2$ results, in very good agreement with the experimental value~\cite{giovannetti07}. 

\subsection{Nickelates of the type RNiO$_3$}
In the examples of type-II MF with the magnetostriction mechanism considered above, the different charges of magnetic ions were determined simply by the structure of the compound: in the previous section we discussed different TM ions such as Co$^{2+}$ and Mn$^{4+}$ in Ca$_3$CoMnO$_6$ or Mn$^{3+}$ and Mn$^{4+}$ in respectively the pyramids and octahedra of RMn$_2$O$_5$. However inequivalent ions can also appear spontaneously, as a site-centered CO, with bonds becoming different due to magnetostriction. Possible examples of this are certain rare earth nickelates RNiO$_3$ (see also~\cite{mostovoy07}).

\begin{figure}
\centerline{
\includegraphics[width=0.65\columnwidth,angle=0]{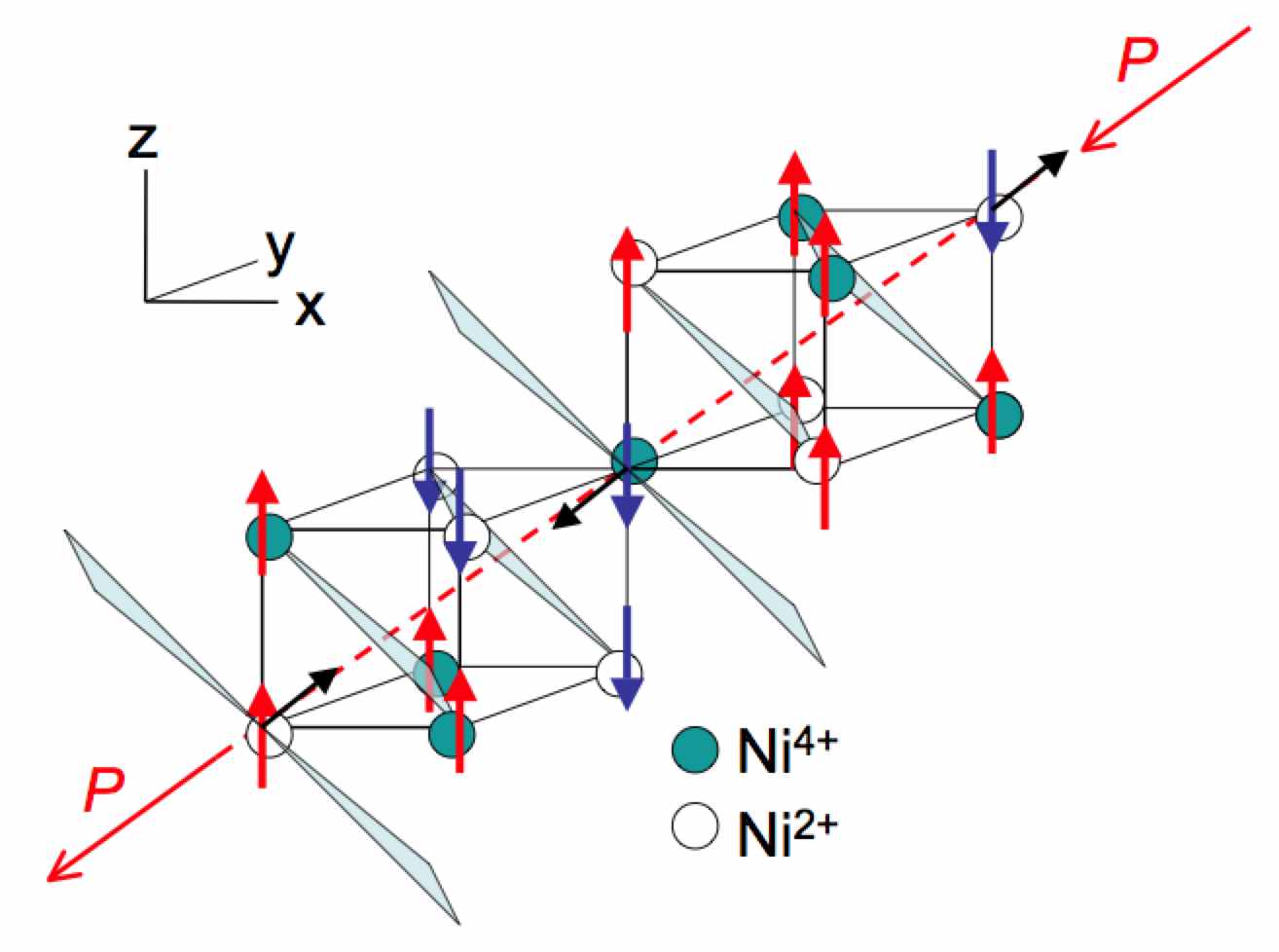}
}
\vspace{0cm}
\caption{Illustration of simultaneous charge disproportionation in Ni$^{2+}$/Ni$^{4+}$ (open and filled circles) and magnetic ordering in RNiO$_3$ (red and blue vertical arrows). The real disproportion will be less ionic and more of the type Ni$^{(3-\delta)+}$/Ni$^{(3+\delta)+}$. Small black arrows indicate the ionic displacements along the body diagonal of the cubes due to magnetostriction, and the direction of the resulting macroscopic electric polarization $P$ is indicated by the red diagonal arrow.}
\label{fig:RNiO3}
\end{figure}

These systems are known to have a spontaneous charge disproportionation (or CO) below their metal-insulator transition \cite{alonso99, mizokawa00}. 
This feature is especially prominent when the rare earth is small (Y, Lu, ...). The disproportionation leads to the formation of a rock-salt-like charge structure: alternation of, formally, Ni$^{2+}$ and Ni$^{4+}$ in consecutive [111] planes the cubic perovskite lattice (for simplicity we ignore orthorhombic tilting in this picture, which can however be important in real materials).

It is still controversial which magnetic structure appears in these systems at temperatures below $T_N<T_{CO}$. One possibility is that it is a sequence of the same [111] planes ordered in $\uparrow \uparrow \downarrow \downarrow$ fashion~\cite{garcia-munoz}. Another proposed structure is more complicated: it has the same superstructure wave vector Q=(1/4, 0, 1/4) (in orthorhombic unit cell), but it has Ni$^{2+}$ and Ni$^{4+}$ spins alternating in different directions, and possibly with non-collinear spins~\cite{alonso99}. The second structure does not lead to FE, at least not in a straight forward fashion. But the first magnetic structure should naturally give a polarization in the same [111] direction: the distances between ferromagnetic
 $ \uparrow \uparrow $ and antiferromagnetic $ \uparrow \downarrow$ planes should not be the same due to different magnetostriction of ferro- and antiferromagnetic bonds, which, for the alternating charges between these [111] layers would immediately give ferroelectricity, as illustrated in Fig.~\ref{fig:RNiO3}.

Note that this mechanism for FE is different from the one proposed in Refs.~\cite{sergienko06,picozzi07} for manganites with E-type antiferromagnetic ordering (which is rather similar to the first magnetic structure of nickelates discussed above, Fig.~\ref{fig:RNiO3}). The mechanism presented in Ref.~\cite{sergienko06} does not require inequivalent charges of TM ions. In addition it would produce a polarization not in the [111] direction, but in the plane perpendicular to it.

\subsection{Magnetic E-type $R$MnO$_3$ Manganites}
\begin{figure}
\centerline{
\includegraphics[width=0.5\columnwidth,angle=90]{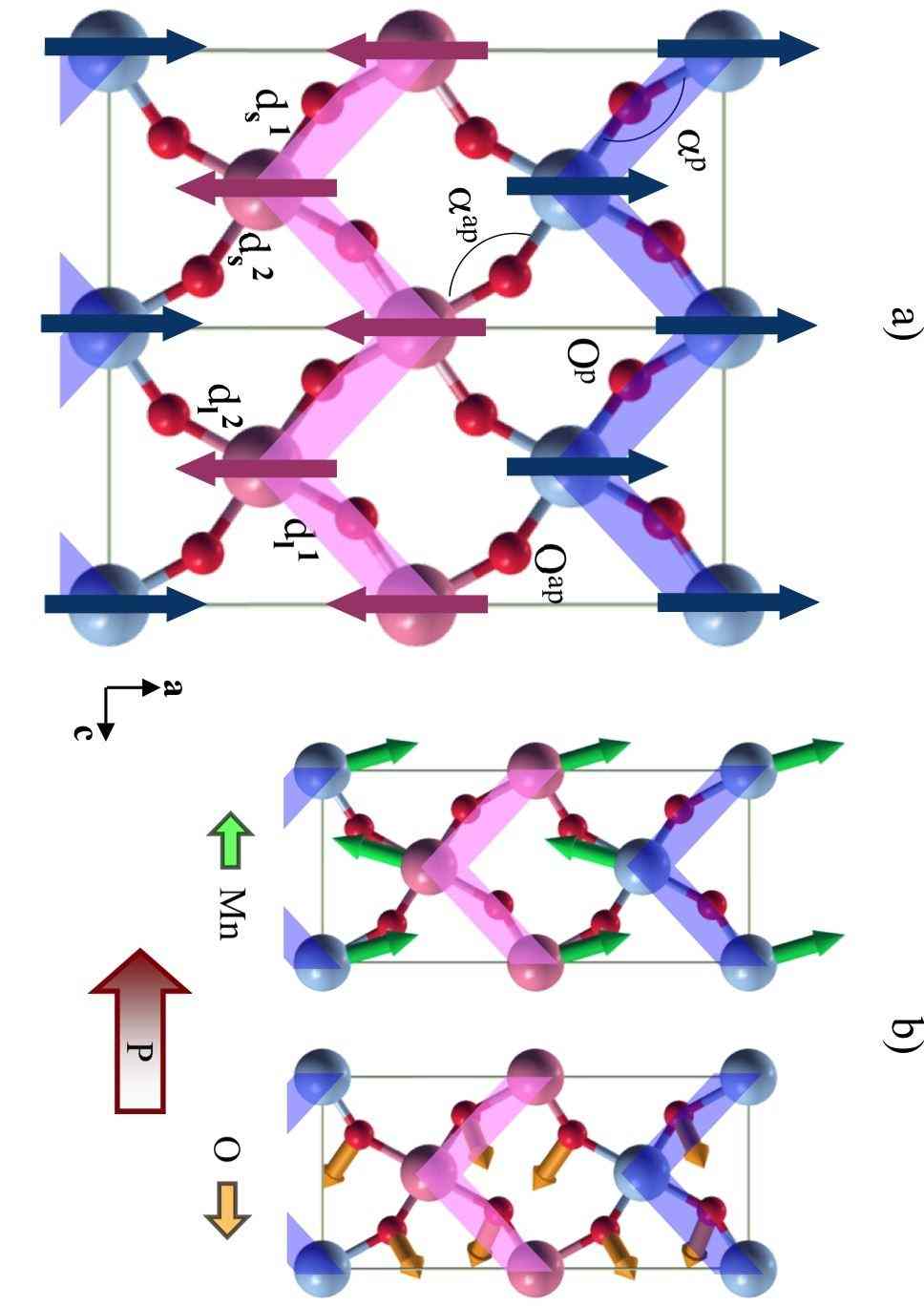}
}
\vspace{0cm}
\caption{
(a) In-plane arrangement of Mn and O atoms in orthorombic manganites with E-type magnetic structure. Arrows denote the direction of spins and AFM-coupled zigzag spin chains are highlighted by shaded areas. The Mn-O-Mn angle for parallel ($\alpha^p$) and antiparallel ($\alpha^{ap}$) Mn spins, large ($d_l^1$ and $d_l^2$)  and small ($d_s^1$ and $d_s^2$) Mn-O bond lengths are indicated. (b) Arrows show the directions of the ionic displacements for Mn (left) and O (right) in the AFM-E phase. The thick arrows at the bottom show the direction of  the resulting displacements of Mn and O sublattices and the polarization P~\cite{picozzi07}.}
\label{fig:silvia}
\end{figure}

In the perovskite manganite family $R$MnO$_3$ the magnetic E-phase was first reported for $R$=Ho~\cite{munoz01}. The E-type magnetic structure of this multiferroic is schematically shown in Fig.~\ref{fig:silvia}. Here we wish to point out that the established picture of multiferroicity in E-type manganites is actually a realization of the same scenario that we have discussed above, of inequivalent bonds and sites combined with striction. Before explaining this in detail, let us retrace our steps for a moment and discuss the E-type materials in a bit more general context. 

As already stated, we can separate type-II multiferroics into to subcategories: those in which ferroelectricity is due to a particular non-collinear (spiral) magnetic structure~\cite{nagaosa,mostovoy07}  -- with as microscopic driving mechanism of the polarization an inverse Dzyaloshinskii effect~\cite{sergienko06} --  and those in which FE is due to magnetostriction. Examples of the second class are e.g. RNiO$_3$ and RMn$_2$O$_5$. These might seem very different systems, but they actually have, as we pointed out, the same physics as discussed in the main part of our paper: inequivalent sites to start with, plus bonds becoming inequivalent due to magnetostriction. 

\begin{figure}
\centerline{
\includegraphics[width=0.8\columnwidth,angle=0]{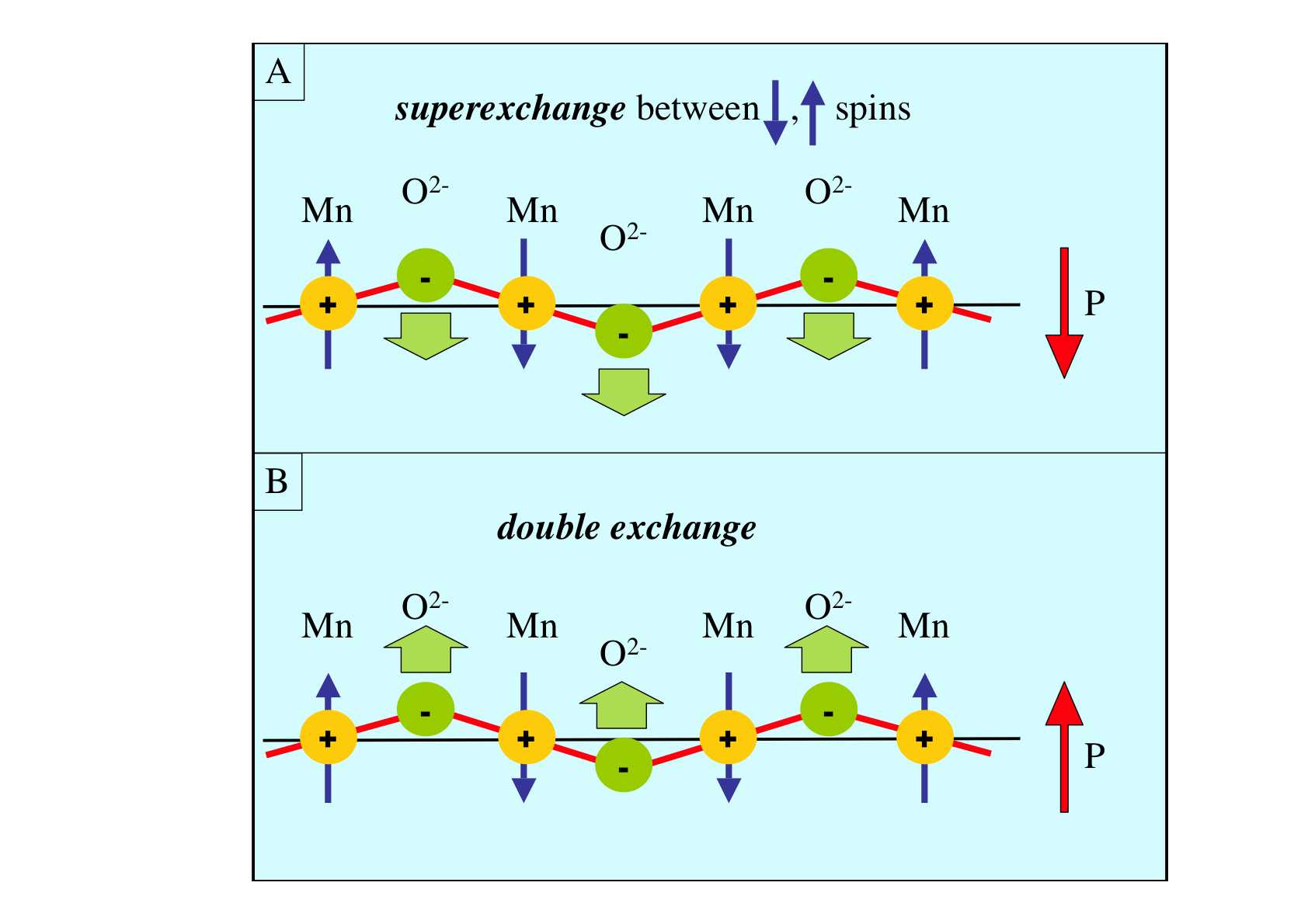}
}
\vspace{0cm}
\caption{
(a) Schematic view of the arrangement of Mn and O atoms in presence of a GdFeO$_3$ distortion. Blue arrows denote the direction of Mn spins in the
"straight" Mn chains of the magnetic E-type ordering, cf. Fig.~\ref{fig:silvia}. Superexchange  drives the Mn-O-Mn angle towards 180 degrees when spin are antiparallel and away from it when the spins are parallel. Green arrows indicate the resulting oxygen displacements; the direction of the ferroelectric polarization is indicated by the red arrow. (b) Double exchange  also causes a ferroelectric polarization, but in the opposite direction. It drives the Mn-O-Mn angle away from 180 degrees when spin are antiparallel and towards it when spins are parallel.}
\label{fig:magnetic_shift}
\end{figure}

In the materials discussed so far, however, one  started out with TM ions with different charge (different valence). The suggested mechanism of FE in the E-type manganites~\cite{sergienko06_2}, supported by ab-initio calculations~\cite{picozzi07}, seems at first glance different: all metal ions have the same valence and charge (Mn$^{3+}$). However in fact the situation is rather similar to the one discussed in the previous section, with positively-charged Mn$^{3+}$ and negatively-charged O$^{2-}$ playing the same role as TM ions with different valence is systems described above. This point is schematically explained in Fig.~\ref{fig:magnetic_shift}, which is a simplified representation of the real crystal and magnetic structure of Fig.~\ref{fig:silvia}. 

As shown in Figs.~\ref{fig:silvia} and \ref{fig:magnetic_shift}, the oxygens are shifted away from their original position at the center of TM-TM bonds, due to the usual tilting of TMO$_6$ octahedra that is present in perovskites (the GdFeO$_3$ distortion). Note that at this point all TM-O-TM bonds are still equivalent. However, in the E-type magnetic structure the spin orientation along the TM-O-TM chains is up-up-down-down. To optimize the exchange in FM and AFM bonds the lattice can distort, which in this case, in contrast to Fig.~\ref{fig:bond-site-cdw}, will be achieved not by longitudinal shifts of the TM ions, but by transverse shifts of the oxygens, modulating the TM-O-TM angles and consequently modifying exchange constants. For localized electrons, according to Goodenough-Kanamori-Anderson rules for superexchange, the strength of a ferromagnetic bond is increased by {\it decreasing} the TM-O-TM angle (thus moving the oxygens in this bond down in Fig.~\ref{fig:magnetic_shift}A). To strengthen an AFM bond, however, the TM-O-TM angle should {\it increase}, approaching 180 degrees. In the magnetic E-phase this requires a shift of respective oxygens in the same direction, see Fig.~\ref{fig:magnetic_shift}A. 

It is interesting to note that in the case of double exchange between TM ions the oxygen would shift in the opposite direction: in this situation the ferromagnetic exchange increases with the TM-O-TM angle stretching to 180 degrees, see Fig.~\ref{fig:magnetic_shift}B. This picture applied to the E-type manganites gives the result indicated in the right panel of Fig.~\ref{fig:silvia}b.

In both cases the net effect is that all negatively-charged ions (oxygens) shift in the same direction, which results in a net electrical polarization perpendicular to the chain. One can show that after such a distortion the gain in exchange energy is linear in displacements $u$, whereas the cost in elastic energy is proportional to $u^2$, i.e. this process will always be favorable, given the crystal and magnetic structure that we discuss here. Clearly this is mechanism  very similar to the one described earlier in this review. It also relies on the presence of ions with different charge (here positive TM ions and negative oxygens), together with bonds which become inequivalent due to a particular (here E-type) magnetic structure. Such bond-bending mechanism is apparently rather effective; the corresponding elastic modulus that counteracts such distortion is usually small. Therefore it can give rise to a rather large polarization~\cite{picozzi07}. 

\subsection{Bilayer manganite (LaSr)$_3$Mn$_2$O$_7$}
One more example of  multiferroicity that is in its nature related to the charge ordering, is observed in the bilayer manganite (LaSr)$_3$Mn$_2$O$_7$~\cite{tokunaga06}. In this paper the evidence for a reorientation of orbital and magnetic stripes below 300 K is presented. After this happened,  the situation becomes again similar to that shown in Fig.~\ref{fig:bond-site-cdw}, with striction plus CO leading to FE. Thus, the mechanism of creation of FE here is similar to the one discussed above, although it remains unclear so far {\it why} stripes rotate by 90 degrees at this transition.

\section{Conclusions}
In conclusion, we have described in this paper a generic mechanism that generates ferroelectricity in systems with charge ordering, which, in particular, can lead to multiferroic behavior in transition metal compounds. This mechanism relies on the simultaneous presence of inequivalent sites and bonds. This can occur spontaneously, due  to coexistence of cite- and bond-centered charge ordering or charge density wave -- e.g. in the case of (PrCa)MnO$_3$. The other possibility is that one of these features is a property of the system itself, where either bonds or sites are inequivalent due to the very chemical or crystal structure of the compound. In case of inequivalent bonds, the sites may become different due to charge ordering; the example of that is organic compounds (TMTTF)$_{2}$X or LuFe$_2$O$_4$.  Such systems may be classified as type-I multiferroics -- the systems in which magnetism and ferroelectricity have different mechanisms and are in principle independent of each other. 

If the sites are different, e.g. have different valence, a bond alternation producing ferroelectricity may occur for example due to magnetostriction. In this case ferroelectricity appears only in a magnetically-ordered state with a particular magnetic structure: these systems are type-II multiferroics. However, in contrast to most other multiferroics of this class, this mechanism does not require relativistic spin-orbit coupling -- generally a small effect in transition metal compounds.

One of the consequences is that in principle charge ordering-induced multiferroicity may lead to a rather large polarization, comparable to that of good ferroelectric like BaTiO$_3$ and much larger than in  most of the type-II (magnetically-driven) multiferroics. The example of LuFe$_2$O$_4$ demonstrates this. On the other hand, how strong the magnetoelectric coupling in this case will be and what will actually determine it,  is still an open question. In any case, this mechanism of multiferroic behavior is certainly interesting and promising. Other systems of this type, besides the ones we have discussed in the present paper, will most probably continue to be discovered. 

\subsection{Acknowledgements}
We are indebted to a great number of theoretical and experimental colleagues for stimulating and fruitful discussions on the topic of charge ordered multiferroics. In particular we would like to thank Marisa Medarde, Paolo Radaelli, Maxim Mostovoy, Serguei Brazovskii, Nicola Spaldin, Silvia Picozzi, Dimitry Efremov and Gianluca Giovannetti. We thank Serguei Brazovski for providing us with Fig.~\ref{fig:phase1d}, Maxim Mostovoy for Fig.~\ref{fig:HoMn2O5} and Silvia Picozzi for Fig.~\ref{fig:silvia}. The work of JvdB is supported by the Stichting voor Fundamenteel Onderzoek der Materie (FOM). The work of D.~Kh. is supported by the Deutsche Forschungsgemeinschaft (DFG) via SFB 608.


\begin{thebibliography}{99}
\bibitem{remark_homogeneous} We have in mind homogeneous systems, not the multicomponent ones.
\end{thebibliography}
\end{document}